\documentclass{kapproc} 

\usepackage{procps} 

\usepackage{psfig}

\upperandlowercase

\kluwerbib 

\def\la{\raisebox{-0.5ex}{$\,\stackrel{<}{\scriptstyle\sim}\,$}}
\def\ga{\raisebox{-0.5ex}{$\,\stackrel{>}{\scriptstyle\sim}\,$}}

\def\teff{\ifmmode T_{\rm eff} \else $T_{\mathrm{eff}}$\fi}

\def\ltsima{$\buildrel<\over\sim$}
\def\lsim{\lower.5ex\hbox{\ltsima}}

\newcommand{\hii}{H~{\sc ii}}
\newcommand{\ha}{\ifmmode {\rm H}\alpha \else H$\alpha$\fi}
\newcommand{\hb}{\ifmmode {\rm H}\beta \else H$\beta$\fi}
\newcommand{\lya}{\ifmmode {\rm Ly}-\alpha \else Ly-$\alpha$\fi}

\newcommand{\heii}{He~{\sc ii}}
\newcommand{\Heiiuv}{He~{\sc ii} $\lambda$1640}
\newcommand{\Heiiopt}{He~{\sc ii} $\lambda$4686}
\newcommand{\qh}{\ifmmode q({\rm H}) \else $q({\rm H})$\fi}
\newcommand{\qhe}{\ifmmode q({\rm He^0}) \else $q({\rm He^0})$\fi}
\newcommand{\qhep}{\ifmmode q({\rm He^+}) \else $q({\rm He^+})$\fi}
\newcommand{\Qh}{\ifmmode Q({\rm H}) \else $Q({\rm H})$\fi}
\newcommand{\Qhe}{\ifmmode Q({\rm He^0}) \else $Q({\rm He^0})$\fi}
\newcommand{\Qhep}{\ifmmode Q({\rm He^+}) \else $Q({\rm He^+})$\fi}
\newcommand{\Qhtwo}{\ifmmode Q({\rm LW}) \else $Q({\rm LW})$\fi}
\newcommand{\Qrathep}{\ifmmode Q({\rm He^+})/Q({\rm H}) \else $Q({\rm He^+})/Q({
\rm H})$\fi}
\newcommand{\Qrathepave}{\ifmmode \bar{Q}({\rm He^+})/\bar{Q}({\rm H}) \else $\b
ar{Q}({\rm He^+})/\bar{Q}({\rm H})$\fi}

\begin{document}

\articletitle[Primeval galaxies]
{Expected properties of 
primeval galaxies and
confrontation with existing observations}

\author{Daniel Schaerer}

\affil{Geneva Observatory, 51, Ch. des Maillettes, CH-1290 Sauverny, Switzerland}
\email{daniel.schaerer@obs.unige.ch}






\vspace*{0.5cm}

\section{Properties of low metallicity and PopIII starbursts
-- predictions and confrontation with existing observations}
Based on existing or new stellar tracks and appropriate non-LTE model
atmospheres we have recently undertaken a systematic study
of the expected restframe EUV, UV to optical properties of starbursts
of zero and higher metallicity
(Schaerer 2002, 2003, hereafter S02, S03).
These studies provide predicted SEDs including stellar and nebular
emission (continuum and lines), detailed ionizing properties,
calibrations for star formation indicators (S03), and also
predicted metal production rates for various elements (S02).
These should be useful for a variety of studies concerning 
high redshift galaxies, cosmological reoinisation studies and others.

\begin{figure*}[htb]
\centerline{\psfig{file=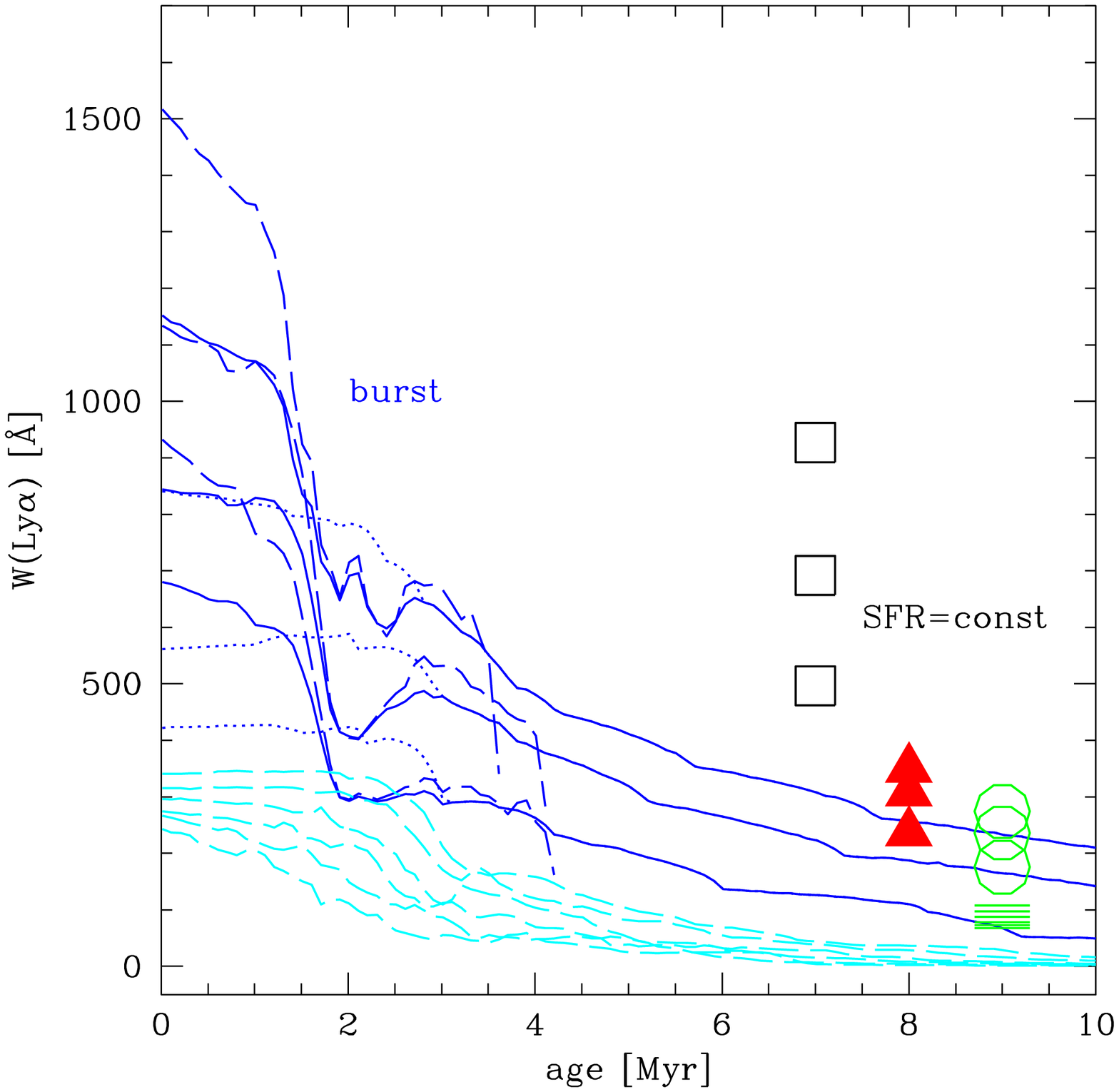,width=7cm}
            \psfig{file=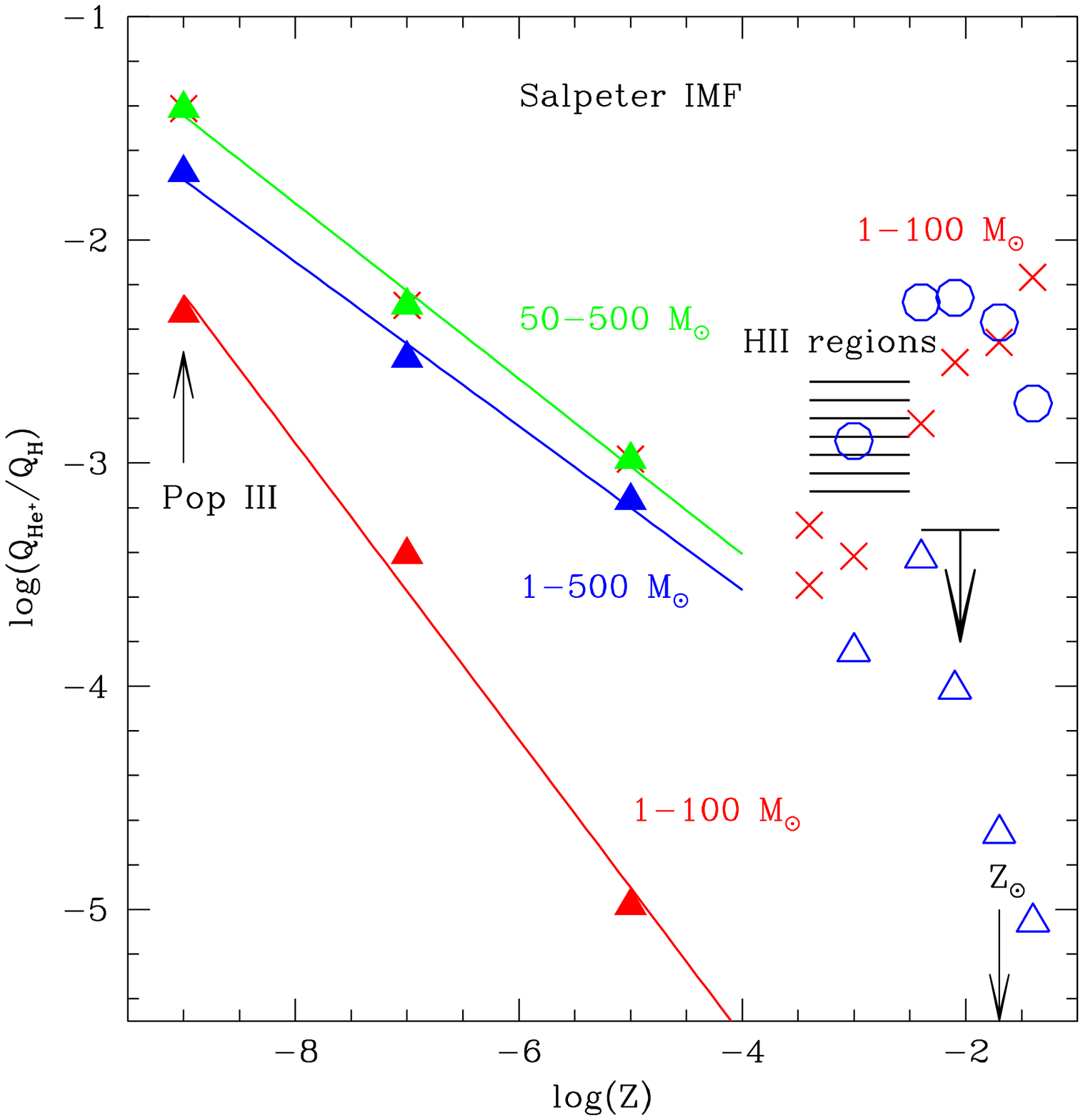,width=7cm}}
\caption{{\bf a--left:}
Predicted \lya\ equivalent widths for bursts of different
metallicities and IMFs.
{\bf b--right:}
Predicted hardness \Qrathep\ as a function of metallicity
for starbursts between PopIII and normal metallicities.
Figures taken from Schaerer (2003) -- see there for a complete
explanation of the figures}
\end{figure*}

Among the main observational characteristics found for PopIII
and metal-poor starbursts are:

{\em 1)} Continuum emission is expected to be dominated by nebular 
(bound-free, free-free and two photon) emission even
in the rest-frame UV domain (S02).
This leads to a flatter intrinsic SED compared to normal galaxies.

{\em 2)} The maximum \lya\ emission (i.e.\ neglecting possible
radiation transfer effects trought the galaxian ISM and IGM and dust)
of integrated stellar populations increases quite strongly with 
decreasing metallicity. This is illustrated in Fig.\ 1a considering
also various cases of the IMF at low metallicity. 

{\em 3)} Strong \heii\ recombination lines (\Heiiuv,
\Heiiopt,...) are a quite unique signature due to hot massive main 
sequence stars of PopIII/very low metallicity
(cf.\ Tumlinson \& Shull 2000). Significant \heii\ emission
is, however, only expected at metallicities below $\la 10^{-5}$ solar (S03).
This is e.g.\ illustrated by plotting the ratio of He$^+$ to H ionizing
photons, \Qrathep, computed for a SF population at equilibrium, i.e.\
for constant SF (Fig.\ 1b).

These findings are also confirmed by the study of Panagia, Stiavelli, 
and collaborators (see these proceedings).

In short, the following three characteristics can be identified
as fairly clear signatures of very metal-poor or PopIII starbursts
(``primeval galaxies''):
Strong \lya\ emission, narrow nebular \heii\ emission,
and very weak or absent metal lines.
We now discuss more precisely these ``criteria'' and confront
them with existing observations.

\subsection{Strong \lya\ emission}
If \lya\ equivalent widths $\ga$ 500 \AA\ are measured,
this would, according to Fig.\ 1a, require stellar populations
with metallicities $Z\la 10^{-5}$ (i.e. [Z/H] $\la -3.3$ in solar
units) and/or an extreme IMF. 
Such large $W(\lya)$ cannot be explained by stellar photoionisation
in galaxies with a Salpeter like IMF and higher metallicities.
Alternative explanations include AGNs.

In fact a good fraction of the \lya\ emitters found by 
the LALA survey seems to show such large equivalent widths (Malhotra
\& Rhoads 2002).
However, it must be remembered that the equivalent widths are
determined from narrowband (NB) and $R$-band imaging, which could
lead to important uncertaintied, as e.g.\ noted by
comparing similar data from NB imaging and spectroscopy of SUBARU
data of Ouchi et al.\ (2003 and private comm.).
Spectroscopic follow-up of 5 LALA sources seem, however,
to confirm their previous equivalent widths measurements from
imaging (Rhoads et al.\ 2003).
In principle it is thus possible that (some of) the strong \lya\ 
emitters of the LALA are such primeval galaxies.
Whether such large numbers of objects at relatively low redshifts
(e.g.\ at $z=4.5$ Malhotra \& Rhoads 2002) would truly be expected
seems surprising (cf.\ Scannapieco et al. 2003).
No doubt detailed follow-up work on these objects will be of
great interest.

\subsection{Nebular \heii\ emission}
As mentioned above the detection of nebular \heii\ emission lines
above a certain level (e.g.\ $W($\Heiiuv$) \ga$ 10 \AA\
or \Heiiuv$/\hb \ga 0.05$, see S03)
would be a very strong case for a very metal-poor if not metal-free
stellar population if shown to originate from stellar photoionisation
(as opposed e.g.\ to AGN activity).

Tumlinson et al.\ (2001) have speculated whether such objects
could already have  been found ``accidentaly'' at $z\la 5$,
confused with \lya\ emitters whose redshift is identified by a single line
measurement.
Their simulations (in rough agreement with our more detailed results)
show that this should then correspond to PopIII objects with 
high star formation rates comparable to those determined for Lyman
break galaxies. 
This may again be questionable on grounds of the 
fairly small probability of existence of PopIII galaxies at
such ``low'' redshifts. However, the current observations do not allow to
prove or disprove their speculation.

During this conference my attention was brought to the discovery
of an unusual $z=3.357$ lensed galaxy showing narrow \heii\ emission
lines (among others) indicative of a fairly high excitation
(see Fosbury et al.\ 2003).
These authors speculate that this might be ionised by a very hard
stellar spectrum possibly due to an extremely metal-poor stellar
cluster. However, this seems quite unlikely for the following reasons.
First numerous metal lines are seen (including in the UV spectrum)
indicating an ISM metallicity of $\sim$ 1/20 solar, and
[O~{\sc iii}] $\lambda$5007/\hb\ $\sim$ 7.5 is typical of metal-poor
\hii\ galaxies. Furthermore, no case is known where the stellar 
metallicity is lower than the nebular one.
This would require quite exotic formation and mixing scenarios,
in contradiction with our current knowledge of SF galaxies in 
the nearby Universe.
Finally, an alternative explanation (obscured AGN) exists capable
of reproducing the observed emission line properties (Binette et al.\ 2003).
Although this peculiar object is certainly of great interest it
appears unlikely that it is related to an extremely metal-poor or even
PopIII starburst.

\subsection{Weak/absent metal lines}
Establishing the weakness or absence of metal lines will ultimately be
important to prove the extremely low or zero metallicity case.
From simple photoionisation modeling Panagia (these proceedings and 2003)
propose e.g.\ that [O~{\sc iii}] $\lambda$5007/\hb\ $\la$ 0.1 (0.01) should
indicate a metallicity $\la 10^{-3} \,\, (10^{-4})$ solar, approximately
corresponding to the expected metallicity of second generation
stars (see also Scannapieco et al.).
Above redshift $z \ga 4$, a measurement of [O~{\sc iii}] $\lambda$5007 will require
sensitive space instruments like those planned for the JWST.
Up to redshift $z \la 9$ such observations should be feasible with the
NIRSpec multi-object spectrograph (e.g.\ Panagia 2003).
Beyond that, explorations will be very difficult and time consuming
as single object spectroscopy (MIRI) will only be available at 
the corresponding wavelengths ($\lambda > 5 \mu m$).

\section{Into the future...}
From the above it is evident that no genuine PopIII object or
extremely metal-poor galaxy ($Z/Z_\odot \la 10^{-3\ldots-4}$) has been 
found so far. Currently the best candidates are the 
high \lya\ equivalent widths objects from the LALA survey whose
properties remain puzzling. However, given their fairly large number
and relatively low redshift ($z \sim 4.5$) it would at first be surprising if
many would truly correspond to this category.

Other indepedent options to search for such ``primeval'' galaxies 
and higher $z$ objects should be explored.
The feasibility of such studies has been addressed e.g.\ in
Schaerer \& Pell\'o (2001) and Pell\'o \& Schaerer (2002).
Possible avenues include among others the use of particular broad-band selection
criteria and ultra-deep near-IR imaging to find new candidates.
Preliminary results from such studies are discussed by Richard et al.\ 
(2003) in these proceedings.
There is little doubt that the great progress made over recent years on 
the exploration of the high redshift Universe will continue and
possibly even truly ``primeval'' galaxies or PopIII objects will be
found in this decade and before the JWST will be available.
We look forward to exciting times!

\end{document}